\documentclass{emulateapj}

\usepackage{graphicx}
\usepackage{gensymb}
\usepackage{epsfig}
\usepackage{epstopdf}
\usepackage{times}
\usepackage[]{natbib}
\usepackage{url}
\usepackage{color}
\usepackage{amssymb,amsmath}

\newcommand{\sci}{{Science}}
\newcommand{\jcph}{{Journal of Comp. Phys.}}

\shorttitle{Helical blowout jets in the Sun: untwisting and propagation of waves.}
\shortauthors{Lee et al.}

\begin{document}

\title{Helical blowout jets in the Sun: untwisting and propagation of waves.}

\author{E. J. Lee,
V. Archontis and A.W. Hood}

\affil{School of Mathematics and Statistics, University of St. Andrews, St. Andrews, KY169SS, UK}

\email{ejl32@st-andrews.ac.uk}


\begin{abstract}
We report on numerical experiment of the recurrent onset of helical ``blowout'' jets in an emerging flux region (EFR).
We find that these jets are running with velocities of $\sim 100-250\,\mathrm{km/s}$ and they transfer a vast amount of
heavy plasma into the outer solar atmosphere. During their emission, they undergo an untwisting motion as a result of
reconnection between the twisted emerging and the non-twisted pre-existing magnetic field in the solar atmosphere. For the 
first time in the context of blowout jets, we provide a direct evidence that their untwisting motion is associated with
the propagation of torsional Alfv\'{e}n waves in the corona.
\end{abstract}


   \keywords{Sun: activity -- Sun: interior --
                Sun: Magnetic fields --Magnetohydrodynamics (MHD) --methods: numerical
               }

\section{Introduction}

Solar jets have been observed at various wavelengths (e.g. H$\alpha$ \citep[]{schmieder95, canfield96}, 
EUV and X-ray \citep[]{shibata92, alex99}). They occur
in EFR \citep[e.g.][]{shibata92}, active regions \citep[e.g.][]{canfield96, shimojo96}
, coronal holes \citep[e.g.][]{wang98, cirtain07} etc.
Recently a dichotomy of jets was suggested by \cite{moore10}.
About two thirds of the observed jets fit the ``standard'' reconnection picture, 
which invokes reconnection between oppositely directed magnetic fields, e.g. an emerging
and a pre-existing magnetic field \citep[e.g.][]{heyvaerts77}. The other one third have been termed ``blowout'' jets, which are
triggered by an eruption. More precisely, it has been observed \citep[e.g.][]{moore10, sterling10, shen12} that the
precursor of a ``blowout'' jet, is a twisted and/or a sheared arch, which often carries a
(small) filament or flux rope within it. When this structure becomes unstable and erupts, it blows out the
envelope field producing an untwisting ejection of cool (e.g. chromospheric) and hot material. This
``blowout'' jet is a broad, curtain-like structure in opposition to the ``standard'' jet, which is
more elongated and it is not commonly associated with an eruption event.

Three-dimensional (3D) simulations on the formation of 
blowout jets, during the emergence and eruption of solar magnetic fields, have been carried out by \cite{arch13},
\cite{moreno13} \& \cite{fang14}.
\cite{arch13} have shown that the interaction between an emerging (twisted) magnetic field and an ambient (non-twisted) magnetic field in the
solar corona can trigger both standard and ``blowout'' jets. Their experiments reproduced some of the observed characteristics of the 
blowout jets, their internal helical structure and their overal curtain-like shape.
\cite{moreno13} studied the recurrent onset of eruptions in an EFR and their possible relationship to the subsequent emission of 
``blowout'' jets.
Using a similar numerical set-up, \cite{fang14} showed that heat conduction leads to an increase of the total mass ejection 
in the corona during the emission of the ``blowout'' jets. In previous simulations, \cite{shibata85} had shown that an unwinding 
jet could be the result of reconnection between a twisted loop and an open flux tube. The unwinding motion, which has also been reported in observations 
\citep[e.g.][]{cirtain07,moore13} has been interpreted as the propagation 
of torsional Alfv\'{e}n waves \citep[see also][]{nishizuka08} releasing the stored twist from a twisted magnetic loop into the open ambient field. However, 
no {\it direct} evidence of {\it propagating} Alfv\'{e}n waves in ``blowout'' jets has been provided so far, either on numerical or 
observational studies. 

Here, we report on the recurrent emission of ``blowout'' jets in an EFR with a sea-serpent configuration. We show that the 
``blowout'' jets are untwisted during their ejection and, for the first time, we provide direct evidence of propagating 
torsional Alfv\'{e}n waves during the emission of the ``blowout'' jets.

\section{The model}
We solve the 3D time-dependent, resistive and compressible MHD equations in 
Cartesian geometry, using the Lare3d code \citep[]{arber01}. 
Explicit (uniform) resistivity of $\eta=10^{-2}$ is included.

The initial atmosphere consists of horizontal and homogenous parallel layers in hydrostatic equilibrium. 
The solar interior is represented by a layer in the range ($-5.4\,\mathrm{Mm} \leq z < 0\,\mathrm{Mm}$), 
which is adiabatically stratified. The photosphere/chromosphere layer lies at 
$0\,\mathrm{Mm}\leq z < 1.9\,\mathrm{Mm}$. The temperature at the photosphere is $5100\,\mathrm{K}$ and it 
increases up to $\approx 3 \times 10^{4}\,\mathrm{K}$ at the top chromosphere. The transition region is located 
at heights $1.9\,\mathrm{Mm}\leq z \leq 2.7\,\mathrm{Mm}$. Above it, there is an isothermal layer 
($\mathcal{O}(1)\,\mathrm{MK}$), which is mimicking the lower solar corona ($2.7\,\mathrm{Mm} < z \leq 57.6\,\mathrm{Mm}$). 
In the solar interior, the magnetic field is a horizontal magnetic flux tube, which is twisted. The flux tube 
is located at $z_{0}=-2.1\,\mathrm{Mm}$ and its axis is parallel to the y-axis at $x=0$. The component of the 
magnetic field along the tube's axis (axial field) is given by
\begin{equation}
B_y=B_0\,\mathrm{exp}(-r^2/R^2), \,\,\,\,\, B_\theta=\alpha\,r\,B_y,
\end{equation}
where the tube's radius is $R=450\,\mathrm{km}$ and $r$ is the radial distance
from the tube's axis ($r^{2} = x^{2} + (z - z_{0})^{2}$). The twist of the fieldlines around the tube's axis
is uniform and it is given by $\alpha=2.2 \times 10^{-3}\,\mathrm{km^{-1}}$. With this twist the tube is marginally
stable to the kink instability.
Initially, we apply a deficit in density along the 
tube's axis, making two segments (i.e. at $x=0$, $y=\pm3.2\,\mathrm{Mm}$) more buoyant than the rest of the tube:
\begin{equation}
\Delta\,\rho=[p_t(r)/p(z)]\,\rho(z)\,\mathrm{exp}\,(-y^2/\lambda^2)\sin^{2}\left(2\pi y/\omega\right).
\end{equation}
Thus, $\Delta\,\rho$ is the difference between the background density and the density inside the flux tube after we apply the
density deficit along its axis. The pressure within the flux tube is $p_t$, $\lambda$ defines half the length of each buoyant 
part of the tube, and $\omega$ is defined as half of the flux tube length.
We use $\lambda=3.6\,\mathrm{Mm}$, $\omega=31.5\,\mathrm{Mm}$ and an initial field strength of $B_{0}=2.4\,\mathrm{kG}$, 
which corresponds to plasma ${\beta}\approx 28$.
The corona is filled with an ambient field which is oblique, and is defined by
\begin{equation}
B_c=B_c(z)\,(0,\cos\theta,\sin\theta),
\end{equation}
where $\theta=80\degree$, and $B_c(z)\approx 3\,\mathrm{G}$ for $z \geq 0.54\,\mathrm{Mm}$, and gradually decreases to 0 
for $z < 0.54 \,\mathrm{Mm}$. The numerical domain
is $[-31.5,31.5] \times [-31.5,31.5] \times[-5.4,57.6]\,\mathrm{Mm}$ in the direction perpendicular to the tube ($x$), along the 
tube's axis ($y$) and vertical ($z$), respectively. The numerical grid is a $420^{3}$ box with periodic boundary 
conditions in $x$ and $y$ direction. At the top, there is an open boundary allowing plasma to flow 
out of the grid. There is a non-penetrating, conducting wall at the bottom boundary.

\section{Results and discussion}
\begin{figure}[t]
\centering
\includegraphics[width=0.9\linewidth]{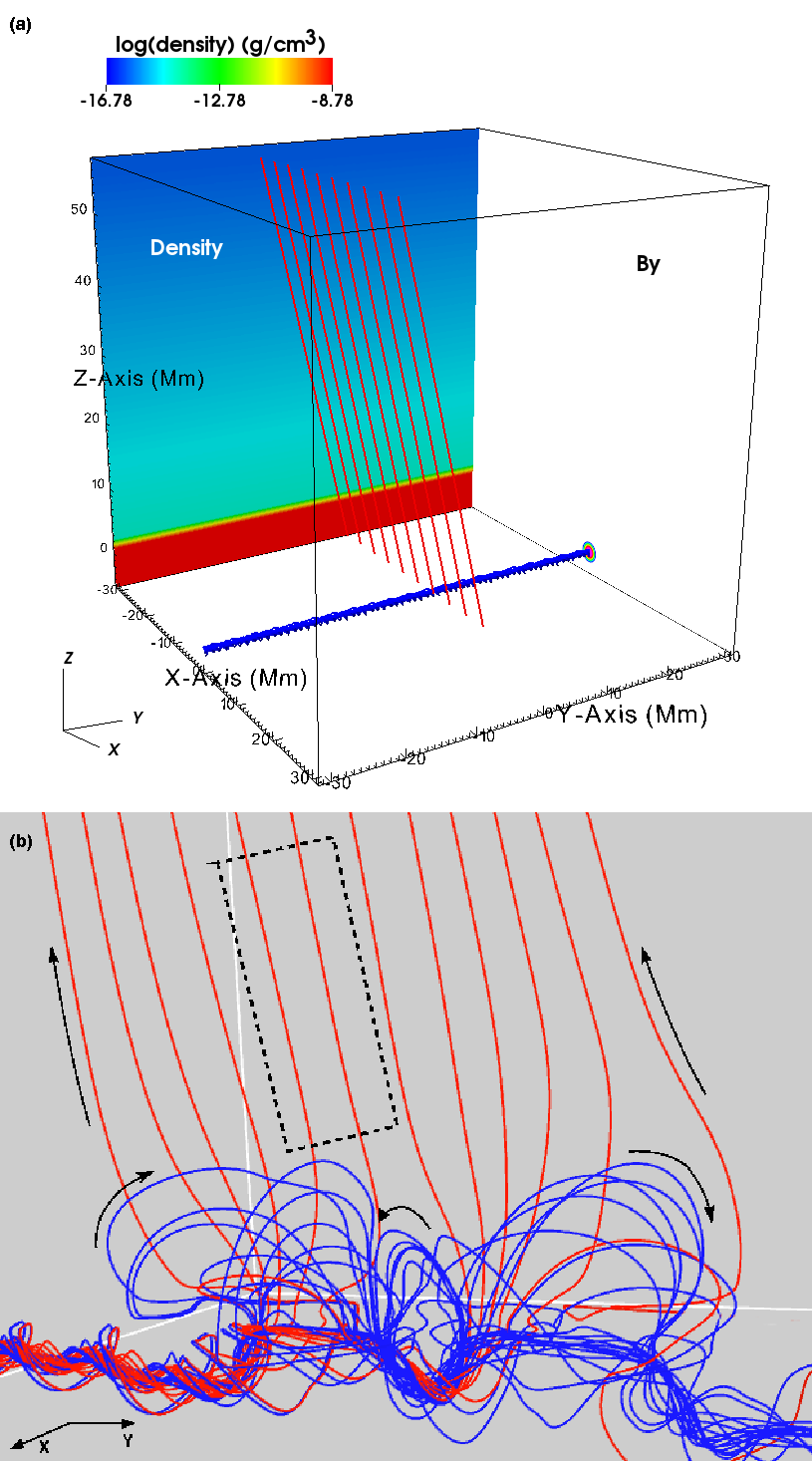}
\caption{({\bf a}) Selected magnetic fieldlines (red) show the orientation of the oblique magnetic field in the corona. The 
twisted magnetic flux tube is shown by the blue fieldlines.
The coloured, vertical y-z plane shows the density distribution of the background 
atmosphere. Isocontours at the x-z plane represent $B_y$. {\bf (b)} A close-up of the fieldline topology ($t=232.8$ min). 
Fieldlines have been traced from the two emerging bipoles (blue) and from the ambient field (red). Arrows show the direction of 
the magnetic field. The (dashed) rectangle indicates the channel along which the first blowout jet is emitted.}
\end{figure}


\begin{figure*}[t]
\centering
\includegraphics[width=0.9\linewidth]{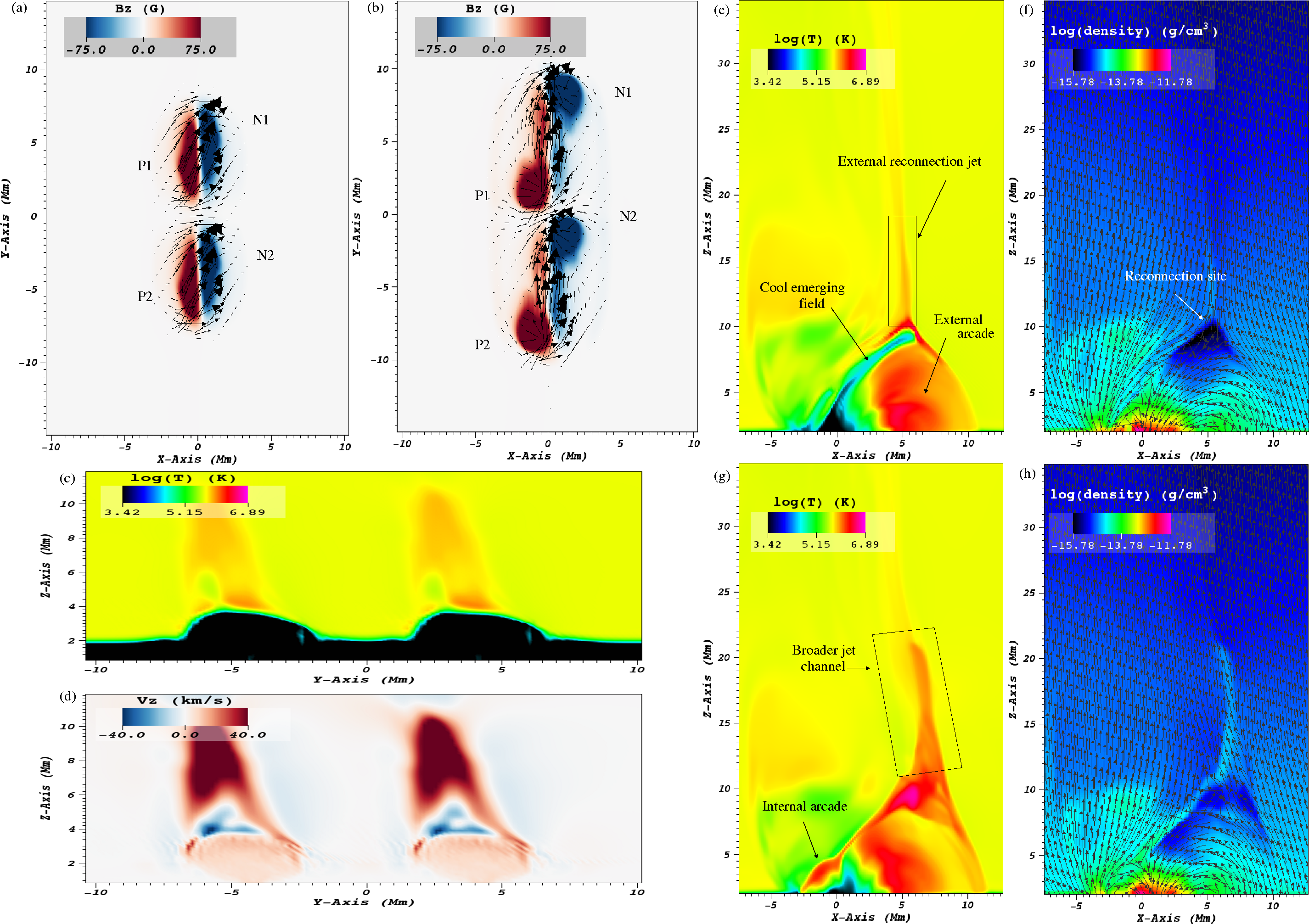}
\caption{The magnetogram during the emergence at the photosphere, showing the two bipolar regions (BR1 
consisting of N1,P1 and BR2 consisting of N2,P2) at two times: {\bf (a)} $t=165.7$ min 
and {\bf (b)} $t=237.1$ min. Temperature {\bf (c)} and 
$v_z$ {\bf (d)} distribution during the emission of the reconnection jets at $t=165.7$ min at $x=0\,\mathrm{Mm}$. 
Temperature {\bf (e)} and density {\bf (f)} distribution during the eruption preceding the first blowout jet at $t=230.0$ min.
Temperature {\bf (g)} and density {\bf (h)} distribution during the
first ``blowout'' jet at $t=232.8$ min. Arrows denote the direction of the magnetic field vector (projected onto the plane).
The vertical slice in {\bf (e-h)} is at $y=-2$ Mm.}
\end{figure*}

\begin{figure}[t]
\centering
\includegraphics[width=0.9\linewidth]{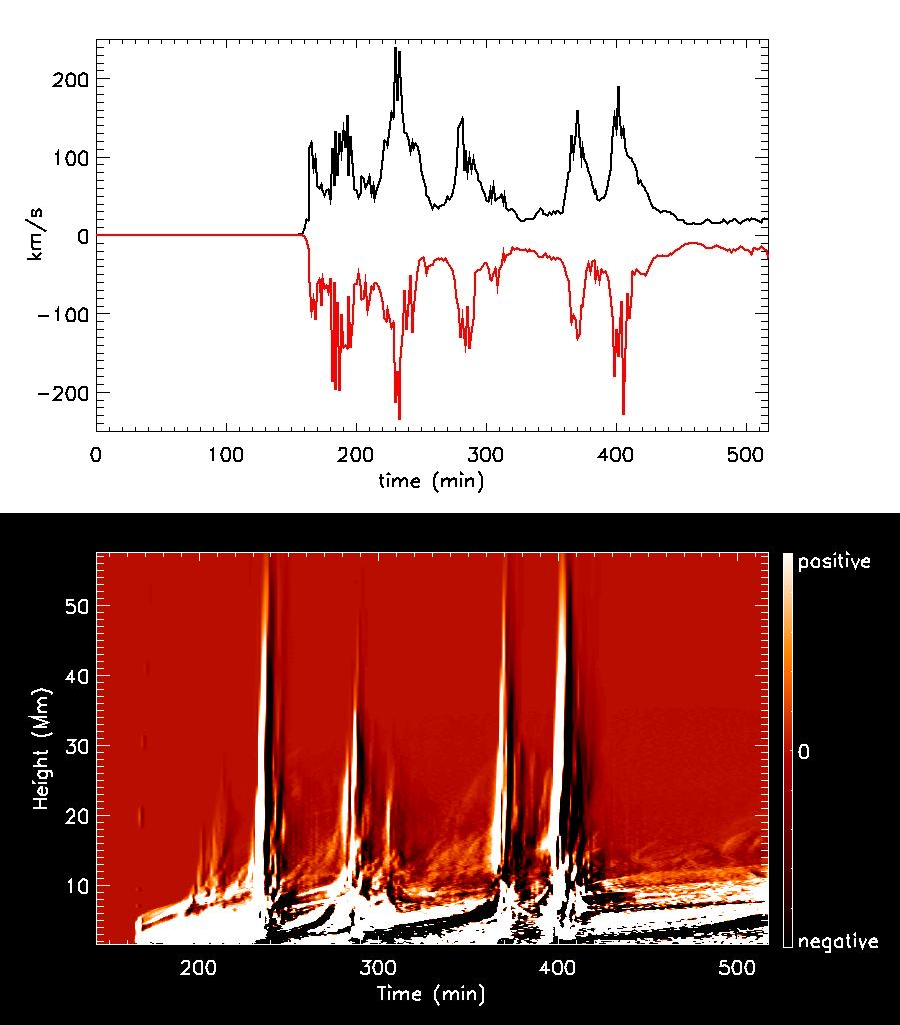}
\caption{{\bf Top:} Temporal evolution of the maximum (black) \& minimum (red) $v_z$ above the photosphere.
{\bf Bottom:} Height-time diagram (running difference) of 
\(\int \! \rho^{2} \, \mathrm{d}x \mathrm{d}y \), where $T > 8\times 10^5\,\mathrm{K}$.} 
\end{figure}

\begin{figure*}[t]
\centering
\includegraphics[width=0.9\linewidth]{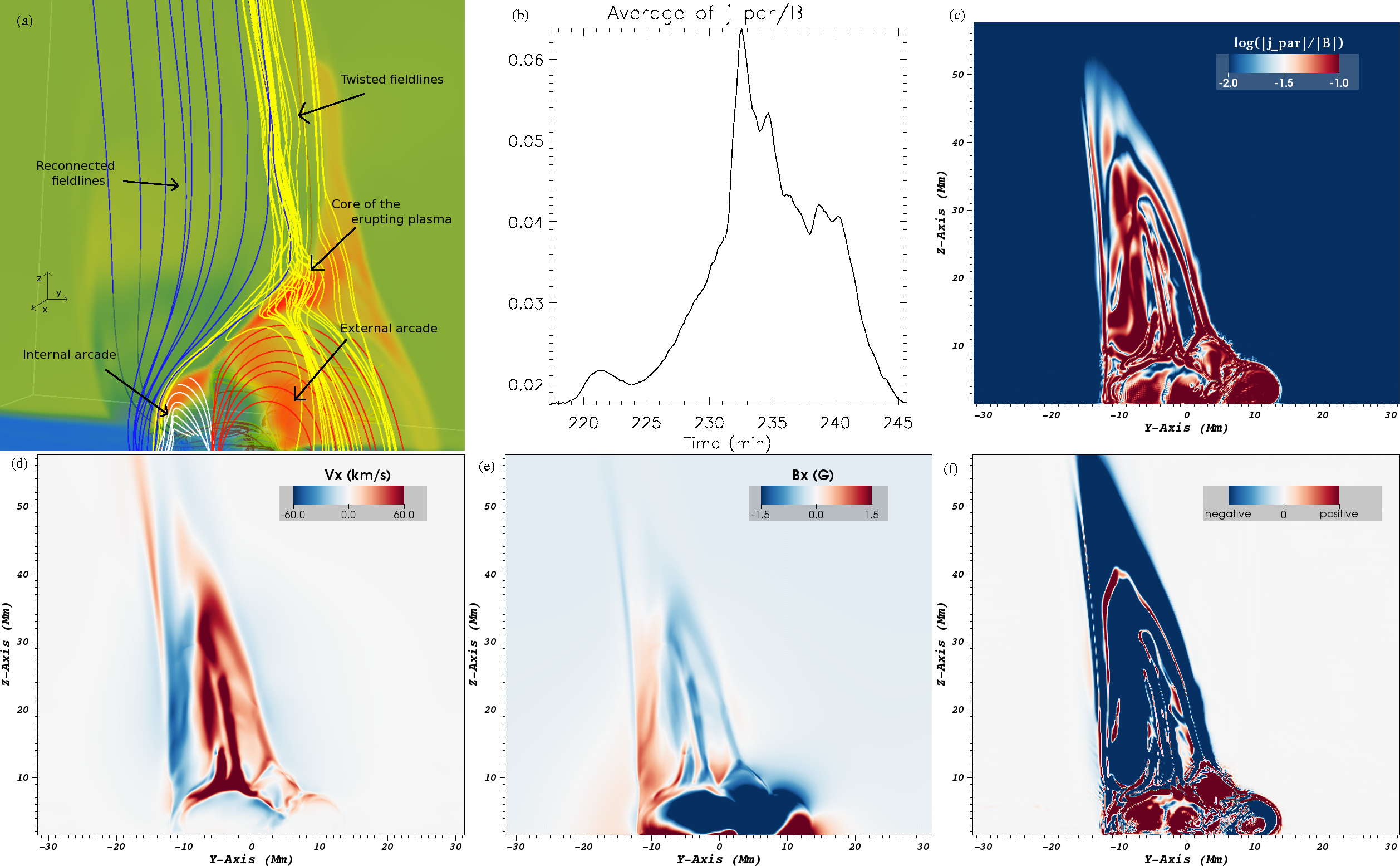}
\caption{{\bf (a)} Visualization of the magnetic
fieldlines, showing the helical nature of the blowout jet (yellow), the reconnected
fieldlines that join the emerging with the ambient field (blue), the internal arcade (white) 
and the external arcade (red). {\bf (b)} The temporal evolution of the average 
$j_{\parallel}/\lvert {\bf B}\rvert$ within the blowout jet.
{\bf (c)}: The $\lvert j_{\parallel}\rvert/\lvert {\bf B}\rvert$ distribution,
{\bf (d)}: $v_x$ distribution, and {\bf (e)}: $B_x$ distribution, at $x=5.4$ Mm, $t=233.4$ min. 
{\bf (f)}: $j_{\parallel}\cdot\omega_{\parallel}$, where $j_{\parallel}$ ($\omega_{\parallel}$) is the
current (vorticity) parallel to the reconnected fieldlines.
}
\end{figure*}

\begin{figure*}[t]
\centering
\includegraphics[width=0.9\linewidth]{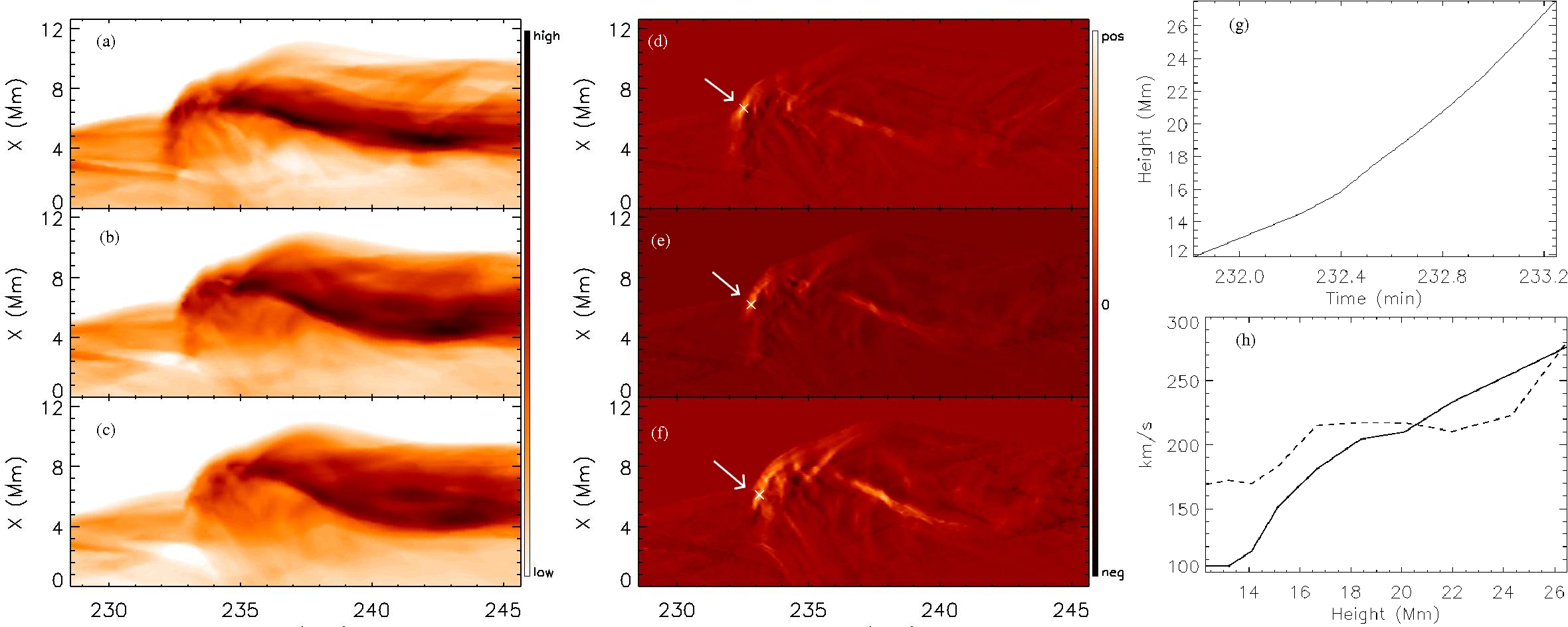}
\caption{The distance-time diagram of \(\int \! \rho^{2} \, \mathrm{d}y \), for 0.6 mK \(< T <\) 1.2 mK,
across the jet at {\bf (a)} 17 Mm, {\bf (b)} 21.5 Mm and {\bf (c)} 25 Mm.
{\bf (d,e,f)} are the running difference of (a,b,c) respectively. {\bf (g)} shows the height-time
profile of the heavy front of the jet, obtained by tracing the first maximum of the running difference (white cross in
(d,e,f)) at various heights. The velocity of this front is shown in panel {\bf (h)}.
The dashed line in {\bf (h)} is the local Alfv\'{e}n speed in the close vicinity of the front of the wave.}
\end{figure*}

In the following, we discuss the results of our simulations showing: (a) the emergence of the sea-serpent flux tube at the photosphere,
(b) the recurrent onset of the jets and (c) the propagation of torsional Alfv\'{e}n waves during the untwisting motion of the blowout jet(s). 
Figure 1(a) shows the initial (i.e. at $t=0$) topology of the magnetic field. 
Figure 1(b) shows the connectivity of the fieldlines during the first blowout jet emission. The 
(blue) fieldlines show (a) the sea-serpent configuration of the emerging field and (b) the twisted fieldlines at the center of the EFR, 
which are the result of reconnection between the emerging loops. The (red) fieldlines are reconnected fieldlines that join the 
ambient with the emerging field. Due to the sea-serpent configuration, the downward tension of the uppermost fieldines (i.e. the 
envelope field) of each emerging loop is released more effectively due to reconnection with ({\it both}) the ambient field and 
the field of the neighbouring emerging loop. Thus, unavoidably, this interaction affects the onset time and dynamics of the EFR's eruptions, compared to 
a ``single emerging loop - ambient field'' reconnection case \citep[e.g.][]{arch13}.

Figure 2(a) shows the magnetogram during the emergence of the field into the photosphere at $t\simeq 165.7\,\mathrm{min}$.
The overplotted arrows represent the direction of the magnetic field vector. Due to the initial density deficit,
the emerging magnetic field at the solar surface forms two bipolar regions (hereafter, BR1 and BR2).
At $t\simeq 237.1\,\mathrm{min}$ (Figure 2(b)), the polarities 
of each BR have moved apart along the $y$-direction.
In previous studies \citep[e.g.][]{arch08}, it has been shown that this movement is followed by shearing along the polarity inversion line (PIL),
and the formation of a new flux rope, which might erupt into the corona. This is also found to happen in the
present simulation, within each BR. Moreover, we find that shearing and reconnection leading to the formation of a flux rope, occurs 
also between the opposite polarities P1 and N2 due to their relative motion.

Before the onset of the eruptions ($t < 200\,\mathrm{min}$), each emerging bipole comes into contact and
eventually reconnects with the ambient magnetic field, giving onset to hot reconnection jets.
Figures 2(c) and 2(d)
show the temperature and $v_{z}$ distribution respectively, at $x=0$ and $t\simeq 165.7\,\mathrm{min}$.
We find that the upward reconnection jets are moving with a velocity of $\sim 40\,\mathrm{km/s}$, reaching
temperatures of up to $1\times 10^6\,\mathrm{K}$ at the reconnection site, and $8 \times 10^5\,\mathrm{K}$
within the jet channels.

Figures 2(e) and 2(f) show the first eruptive event within the EFR. The eruption of the cool plasma starts within BR2.
Figure 2(e) shows the temperature and figure 2(f) shows the density distribution at $y\simeq -2\,\mathrm{Mm}$ and $t=230\,\mathrm{min}$.
The rise of the erupting material induces inflow towards the interface, where a current layer has built up, between the 
envelope and the ambient field. This leads to more external reconnection between the two magnetic flux systems and the
onset of a hot and fast external reconnection jet. The plasma at the interface is heated up to $3.5\times 10^6\,\mathrm{K}$.
A side-effect of the external reconnection is the formation of a hot external arcade with temperature
up to $5.5\times 10^6\,\mathrm{K}$.

At $t\simeq 232.8\,\mathrm{min}$ (Figures 2(g), 2(h)), the eruption blows out the envelope field and the dense erupting material is emitted along the
reconnected fieldlines of the oblique ambient field. Now the channel of the blowout jet and the external arcade are much
wider. Cool ($5\times 10^5\,\mathrm{K}$) and hot ($1\times 10^6\,\mathrm{K}$) plasmas are ejected along the 
blowout jet's channel and heating (due to internal reconnection) is produced
underneath the erupting plasma, where an internal arcade of temperature up to $4\times 10^6\,\mathrm{K}$ is formed.

The top panel in Figure 3 shows that there are six events during which $v_{z}$ varies in the range $\sim 100-250\,\mathrm{km/s}$. 
Each event consists of several peaks in $v_{z}$. The fact
that the peaks of minimum and maximum $v_{z}$ occur at approximately the same time suggests the existence of bidirectional flows 
from several sites within the EFR.
We find that the first peak ($t\approx 165\,\mathrm{min}$) corresponds to the jets caused by reconnection between the emerging and
the ambient magnetic field. 
The second event ($t\approx 185-210\,\mathrm{min},v_{z}\sim150\,\mathrm{km/s}$) 
is the composite effect of an eruption, which starts from the area in between BR1 and BR2, 
and a reconnection jet that follows the eruption. The jet is initiated due to the restructuring of the nearby magnetic
field due to the eruption (in a similar manner to the external reconnection jet in Figure 2e). 
However, in this occasion, the erupting plasma becomes confined by the local ambient field and does not evolve 
into a profound blowout jet that could reach the higher solar atmosphere. Instead, it is moving sideways, from the 
center of the EFR towards the BR1 where the total pressure is less.

The other four events correspond to the emission of ``blowout'' jets driven by eruptions that 
emanate from within the BRs. These jets blast heavy material towards the outer solar atmosphere. 
The bottom panel in Figure 3 shows the running difference of the integrated plasma density
\(I=\int \! \rho^{2} \, \mathrm{d}x \mathrm{d}y \) for temperatures above $8 \times 10^5 \mathrm{K}$. 
It is found that there are four marked events during which dense plasma is brought into the higher atmosphere. 
The steep vertical slope at the end of each event indicates that most of the ejected plasma is 
transported upwards. The dark regions after
the brightening peaks indicate that part of the erupting plasma undergoes gravitational
draining. Notice that the 
time-period between the onset of the blowout jets is not so ``quiet''. There are various, less striking events, 
which do not dump enough mass and energy into the outer corona. Such events (e.g. at $t\approx 190-210\,\mathrm{min}$, 
$t\approx 310\,\mathrm{min}$, etc.) are small confined eruptions and reconnection coronal jets.
By comparing the two panels in Figure 3, we find a good temporal correlation
between the four ``blowout'' jets and the emission of bidirectional flows, which occur after $t\approx 220\,\mathrm{min}$. 
This is a direct evidence that the ``blowout'' jets are closely associated with reconnection, which is driven by the 
eruption of cool material from the low atmosphere.

The erupting core of the blowout jets is a twisted flux tube. 
Figure 4(a) illustrates the magnetic fieldlines within and around the first blowout jet
\citep[see also][]{arch13}. Reconnection of the erupting (twisted) field with the open (non-twisted) ambient field 
would most likely relax the twist and lead to an untwisting motion of the helical blowout jet. A measurement of the 
twist is given by $j_{\parallel}/{\bf B}$, where $j_{\parallel}$ is the parallel component of 
the current along the reconnected fieldlines of the blowout jet. Figure 4(b) shows the temporal evolution of the 
average $\lvert j_{\parallel} \rvert/{\bf B}$ along the channel of the blowout jet. 
We calculate this for heights above $z\approx 9\,\mathrm{Mm}$,     
which is the approximate horizontal boundary between the envelope field of the emerging flux and the lower part of
the blowout jet. The increase of the 
twist (up to $t=235 \,\mathrm{min}$) is due to the eruption of the flux rope and the subsequent decrease indicates the untwisting 
motion of the jet. Figure 4(c) shows the distribution of $\lvert j_{\parallel} \rvert/{\bf B}$ at the 
vertical slice with $x=5.4\,\mathrm{Mm}$. It is confirmed that the blowout jet possess considerable twist along its main stream 
(e.g. in the range $z=10-40\,\mathrm{Mm}$).
Notice that $\lvert j_{\parallel} \rvert/{\bf B}$ is also very strong (as expected) in the twisted 
emerging field underneath the blowout jet (e.g. the area with strong $B_{x}$ at $z<7\,\mathrm{Mm}$).

To show the untwisting motion of the jet, we plot the $v_{x}$ and $B_{x}$ distribution (Figures 4(d) 
and 4(e) respectively). There is positive $v_{x}$ (pointing 
out of the plane) at the right-hand side of the jet, and negative $v_{x}$ (pointing into the plane) at the left-hand side.
This corresponds to a magnetized plasma motion where $B_{x}$ (i.e. the transverse component of the magnetic field)
is pointing in the opposite direction to $v_{x}$ and the associated vorticity (i.e. \(\overrightarrow{\omega}=\nabla\times \overrightarrow{v}\)) is 
pointing downward.
This is a first indication that the jet undergoes an untwisting motion. More evidence 
is found by studying the distribution of current and vorticity within the jet. 

Figure 4(f) shows the product of $j_{\parallel}$ with the vorticity $\omega_{\parallel}$.
Within the jet channel, this product is mostly negative indicating that the current and vorticity are pointing in different 
directions. In fact, we have found (not shown in this figure) that $\omega_{\parallel}$ is negative due to the direction of the 
flow (and it is pointing downwards), while $j_{\parallel}$ is positive due to the direction of the twist of the fieldlines 
(pointing upwards). All the above support the scenario of the untwisting motion of the blowout jet 
during its emission. It is likely that the process of untwisting of the erupting plasma in our experiment 
is similar to the sweeping-magnetic-twist mechanism for the acceleration of solar jets \citep{shibata85}. 
In the present simulation, it is apparent that the twist stored in the 
erupting flux is released along the direction of the open, reconnected
field lines. Also, the direction of the untwisting motion (e.g. as shown in Figure 4)
is consistent with the relative orientation of the two field components
(i.e. the erupting and the open field) that reconnect. Thus, it is more likely that the untwisting 
is related to the propagation of the twist along the reconnected fieldlines 
instead of the twist itself.

To study the plasma motion of the jet, in relation to the direction of the ambient field, we calculate 
\(I_1=\int \! \rho^{2} \, \mathrm{d}y \) at various heights, from $z\approx 12\,\mathrm{Mm}$ to $z\approx 27.5\,\mathrm{Mm}$.
More precisely, we take a horizontal slit at each height
and we plot the above quantity, $I_1$, during the evolution of the system. 
The left panel in Figure 5 shows $I_1$ at $z=17\,\mathrm{Mm}$, $z=21.5\,\mathrm{Mm}$ and $z=25\,\mathrm{Mm}$.
We trace plasma of temperature between $6\times10^5\,\mathrm{K}$ and $1.2\times10^6\,\mathrm{K}$.
We find that firstly the plasma jet is moving towards the positive x direction, then to the negative x direction,
and then back towards the positive x direction. This indicates that the jet undergoes an oscillatory motion during its emission.
A similar result, in the context of helical jets, has been reported by \cite{Liu09}.
The transverse oscillation velocity amplitude of the jet is $\approx 7\,\mathrm{km\,s^{-1}}$.
Figure 5 (a-c) shows that the oscillation propagates from low to larger heights.
Since the jet is considered to be
emitted along the ambient magnetic field, this is an evidence of a wave propagating towards
the outer solar atmosphere.

To measure the velocity of the propagating wave, we use the height-time diagrams in panels (a)-(c) and 
we plot the running-difference of each panel, as shown in panels (d)-(f). 
Then, we find the first (i.e. along time) local maximum in each running-difference plot, which corresponds to the
first profound increment of plasma density at that height. It is likely that this point is located at the 
front of the heavy plasma distribution along the jet.
The height-time profile of this front is then plotted in Figure 5(g) and by calculating
the gradient, we obtain the propagating speed (Figure 5(h)). 
The latter is comparable to the local Alfv\'{e}n speed, which
suggests that we are witnessing the propagation of a torsional Alfv\'{e}n wave during the untwisting motion of the
blowout jet. We find that the above result is generic: all the blowout jets in our experiment show transverse oscillatory motions and 
encompass propagating Alfv\'{e}n waves along their broad outflow streams.

\subsection{Discussion}
The initial ambient open field in our experiment has a field strength of approximately $3\,\mathrm{G}$. The density in the 
high corona is approximately $10^{-16}\,\mathrm{g\,cm^{-3}}$. Therefore, our simulation is mimicking the emission of jets 
in a coronal hole environment. We have calculated that in every blowout jet emission, a considerable amount of 
Poynting energy flux ($\mathcal{O}(10^{5})\,\mathrm{erg\,s^{-1}\,cm^{-2}}$) leaves the high corona. Also the mass deposition in the corona 
increases by a factor of 2-4 during the blowout ejecta.
It is likely that the torsional Alfv\'{e}n waves transport the emitted flux from the low corona to the outer atmosphere 
and load with mass the open field. This implies that blowout jets may play a significant role in driving the solar wind.
\\

The simulations were performed on the STFC and SRIF funded UKMHD cluster, at the University of St Andrews.
VA acknowledges support by the Royal Society.
We would like to thank the anonymous Referee for comments and suggestions on how to improve the manuscript.

\bibliographystyle{apj}

\clearpage

\end{document}